\documentclass{asmeconf}

\usepackage[utf8]{inputenc} 
\usepackage{graphicx}       
\usepackage{amsmath}        
\usepackage{bm}             
\usepackage{hyperref}
\usepackage{url}
\usepackage{float}

\ConfName{Proceedings of the ASME 2025\linebreak International Design Engineering Technical Conferences and \linebreak Computers and Information in Engineering Conference
}
\ConfAcronym{IDETC/CIE2025}
\ConfDate{August 17-20, 2025}
\ConfCity{Anaheim, California}
\PaperNo{IDETC/CIE2025-166080}

\title{Lowering Barriers to CAD Adoption: A Comparative Study of Augmented Reality-Based CAD (AR-CAD) and a Traditional CAD tool}

\SetAuthors{%
 Muhammad Talha\affil{1},
 Abdullah Mohiuddin\affil{1},
 Sehrish Javed\affil{2},
 Ahmed Jawad Qureshi\affil{1}\CorrespondingAuthor{talha4@ualberta.ca, ajquresh@ualberta.ca}
}

\SetAffiliation{1}{Department of Mechanical Engineering, University of Alberta, Edmonton, AB, Canada}
\SetAffiliation{2}{Department of Biomedical Engineering, University of Alberta, Edmonton, AB, Canada}

\begin{document}

\maketitle

\begin{abstract}
The paper presents a comparative user study between an Augmented Reality-based Computer-Aided Design (AR-CAD) system and a traditional computer-based CAD modeling software, SolidWorks. Twenty participants of varying skill levels performed 3D modeling tasks using both systems. The results showed that while the average task completion time is comparable for both groups, novice designers had a higher completion rate in AR-CAD than in the traditional CAD interface, and experienced designers had a similar completion rate in both systems. A statistical comparison of task completion rate, time, and NASA Task Load Index (TLX) showed that AR-CAD slightly reduced cognitive load while favoring a high task completion rate. Higher scores on the System Usability Scale (SUS) by novices indicated that AR-CAD was superior and worthwhile for reducing barriers to entering CAD. In contrast, the Traditional CAD interface was favored by experienced users for its advanced capabilities, while many viewed AR-CAD as a valid means for rapid concept development, education, and an initial critique of designs. This opens up the need for future research on the needed refinement of AR-CAD with a focus on high-precision input tools and its evaluation of complex design processes. This research highlights the potential for immersive interfaces to enhance design practice, bridging the gap between novice and experienced CAD users.

\end{abstract}

\section{Introduction}
In recent years, the confluence of 3D modeling with augmented reality (AR), virtual reality (VR), and mixed reality (MR) technologies has marked a paradigm shift in product design and development methodologies \cite{Billinghurst2005,Wang2006,Funkhouser2013}. Computer-Aided Design (CAD) software---traditionally run on desktop systems using 2D screens and a mouse/keyboard interface---serves as the backbone of modern engineering processes. Dassault SolidWorks, for example, is well-known for its robust 3D modeling and precision, yet emerging AR/VR interfaces promise richer, more intuitive 3D interactions \cite{Lee2005,Bruno2010,Zhang2025}.

Over the past decade, AR has shown particular promise in enhancing manufacturing and product design \cite{Harlan2023,Khan2022,Serdar2013}. Conventional CAD platforms (e.g., SolidWorks, CATIA, Creo) often restrict designers to indirect 3D manipulation with limited spatial awareness \cite{Wang2006}. By contrast, AR-based CAD systems overlay virtual models onto the user’s real-world environment, potentially enabling direct interaction, natural 3D visualization, and collaborative reviews \cite{Billinghurst2005,Lee2005,Zhang2025}. Multiple studies indicate that AR can improve spatial cognition, accelerate geometric conflict detection, and foster dynamic ideation \cite{Wang2006,Funkhouser2013,Yang2019}. Nevertheless, it remains unclear whether AR-based modeling can match or surpass conventional CAD workflows in efficiency and precision, given hurdles such as ergonomic discomfort, limited field-of-view, and relatively immature 3D interaction methods \cite{Harlan2023,Khan2022}.

In industry, leading CAD vendors like PTC (Creo AR Experience) and Dassault Systèmes (eDrawings for AR) have mostly offered AR for design review, but not for full-scale modeling \cite{PTC_AR}. The key motivation for this work is to understand precisely when and how AR-based tools integrate with professional workflows or lower barriers for novices \cite{Billinghurst2005,Harlan2023,Zhang2025}.

This paper addresses these open questions with a comparative user study of AR-CAD versus SolidWorks. We explore differences in modeling efficiency, accuracy, cognitive load, and user comfort, with the ultimate goal of clarifying whether AR-based CAD solutions can coexist with or complement desktop-based systems in practical product development settings.

\subsection{Research Questions}
We adopt a mixed-method evaluation to address:

\noindent
\textbf{RQ1:} Can users design a 3D model in AR-CAD as effectively as in a traditional CAD environment such as SolidWorks?

\noindent
\textbf{RQ2:} How do modeling times (including complex design) differ between AR-CAD and traditional CAD interface?

\noindent
\textbf{RQ3:} How do users perceive usability and comfort in AR-CAD compared to traditional CAD interface?



We hypothesize that AR-CAD reduces barriers for novices and allows a more engaging user experience, while experienced users find traditional CAD tools like SolidWorks more efficient when working on precision and advanced features. We also aim to see whether AR-CAD stands as a potential alternative within a professional design workflow or if it serves the purpose of being a complementary tool.

\section{Related Work}
\subsection{Augmented Reality in CAD}
The incorporation of AR into CAD systems has emerged ever more frequently in research related to intuitive design environments \cite{Billinghurst2005,Lee2005,Zhang2025}. Billinghurst et al. \cite{Billinghurst2005} proposed design principles for AR interfaces, which directly manipulate 3D holographic models whereas Wang and Dunston \cite{Wang2006} showed that the AR "assistant viewer" was significantly more helpful in speeding the detection of spatial conflicts in assemblies than conventional views in CAD. Likewise, Januszka and Moczulski \cite{Januszka2006} presented an AR+CAD system allowing designers to visualize CAD models as holograms within their workspace, thus better grasping complex geometry. The results further supported the idea that placing virtual objects on real scenes provides more intuitive insights into designs \cite{Januszka2006}.

Later prototypes investigated tangible design interactions. For example, Lee and Park \cite{Lee2005} introduced "Augmented Foam," whereby a foam block serves as a tangible interface onto which 3D CAD geometry is projected for direct sculpting; conversely, Bruno et al.\cite{Bruno2010} studied product-behavior simulation within mixed reality, showing ways of visualizing and manipulating mechanical components against real space. Zhang and Wang \cite{Zhang2025} lowered the entry barrier for novices even further with RealityCraft: an in situ AR modeling interface that engages users more than traditional desktop CAD tools.

However, these tools are few and far between in the commercial world. Most commercial AR add-ons are aimed at supporting design review rather than end-to-end modeling \cite{Harlan2023,Funkhouser2013,PTC_AR,Khan2022}. Major platforms, including SolidWorks, eDrawings, and Creo AR Experience, allow viewing or annotating AR models on mobile devices but they do not offer a complete modeling environment, which ultimately requires transitioning back to a traditional CAD environment.

\subsection{AR in Manufacturing and Prototyping}
Apart from design tasks, AR has increasingly important applications in manufacturing, bridging the virtual CAD with the real-world context. The following subsections briefly highlight some key applications.

\subsubsection{Rapid Prototyping and Visualization:}
Augmented reality superimposes digital models over physical backdrops, thus reducing the need for physical prototypes and accelerating the iteration of products \cite{Harlan2023}. An example is Boeing, which used AR to create many design variations without creating physical prototypes \cite{Harlan2023}. Additional studies prove that putting 3D models in place is a more natural way to reveal ergonomic or clearance problems than viewing them on 2D screens \cite{Wang2006,Funkhouser2013}. Januszka and Moczulski also noted how full-scale holographic views can expedite design decisions by giving stakeholders an immediate sense of size and form \cite{Januszka2006}.
\subsubsection{Assembly Guidance:}
The assembly steps of AR-based application systems present guidelines, part identifiers, and aligning markers in front of a technician's view. Yang et al. \cite{Yang2019} found that such systems significantly reduced assembly time and errors compared to paper-based instructions. Technicians at Boeing saw a reduction in mistakes when assembling wiring harnesses with the aid of AR overlays \cite{Harlan2023}. This is especially valuable for a complex or large-scale assembly part because it reduces the cognitive burden of referencing 2D manuals. 
\subsubsection{Training and Skills Development:}
AR-based training modules can standardize procedures, shorten onboarding, and improve worker confidence, particularly for complex processes \cite{Harlan2023,Khan2022,Serdar2013}. By offering real-time overlays that highlight part usage, tool operations, and common pitfalls, AR enables hands-on learning without extensive classroom sessions. Industry reports also show that AR can reduce operator errors and accelerate up-skilling, underscoring its potential as a versatile training aid.

AR training modules are time-efficient and make onboarding procedures more uniform and shorter while increasing employee confidence in end-users, especially for complex processes \cite{Harlan2023,Khan2022,Serdar2013}. Real-time overlays on parts, tool operations, and common pitfalls enable hands-on experience without long hours of lectures. Industry reports also recommend AR as an excellent tool for effective upskilling because it reduces operator errors and provides timely on-the-job training.

\subsection{Comparative Evaluations: AR vs. Traditional CAD}
Comparative user studies showcase AR's significant strengths—especially spatial understanding and conflict detection—while acknowledging limitations in precision and ergonomics for prolonged usage \cite{Harlan2023,Wang2006,Funkhouser2013,Khan2022}. Wang and Dunston's very early work quantified time savings in interference detection aided by an AR-enhanced CAD viewer \cite{Wang2006}. Similarly, Lee et al.\cite{Lee2020} compared AR with a 2D screen and VR in architecture reviews. They reported that AR could give a more realistic contextual visualization, but physical fatigue typically emerged due to the extended use of the headset. 

\subsubsection{AR as Complement, Not Replacement:}
While an immersion tool, AR does not usually have properly constructed parametric modeling or assembly management tools already, the forte of other software like SolidWorks \cite{Harlan2023,Funkhouser2013}. As such, designers often adopt AR for conceptual exploration or design review, then revert to desktop CAD applications for detailed modeling or extended tasks. According to Januszka and Moczulski, AR inspections offered excellent spatial cognition. However, when high precision was required, most designers remained loyal to conventional CAD—a clear testimony to AR being a complement rather than a substitute for desktop systems \cite{Januszka2006}. 

\subsubsection{Hybrid AR Workstations:}
Seeking the best of both worlds, Harlan et al. \cite{Harlan2023} proposed “hybrid AR workstations,” merging AR’s intuitive visualization with the parametric power of desktop CAD. Similarly, more recent approaches integrate live updates between a designer’s monitor and a holographic display, enabling fluid transitions between high-precision modeling tasks and direct 3D inspection \cite{Harlan2022}. Although such hybrid solutions demand robust geometry synchronization and carefully designed user interfaces, they illustrate the future potential of AR in mainstream CAD workflows.
In search of the best of both worlds, Harlan et al. \cite{Harlan2023} touted "hybrid AR workstations," combining AR's intuitive visualization with desktop CAD parametric power. Likewise, recent approaches are now built around live updates between a designer's monitor and a holographic display for fluid transitions between high-precision modeling tasks and direct 3D inspections \cite{Harlan2022}. Geometrical synchronization and interface design are focal points for hybrid setups geared toward AR's prosperous future in mainstream CAD workflows. 

\subsection{Additional AR Interfaces}
\subsubsection{Tangible AR Interfaces:}
Tangible AR interfaces extend tangible user interfaces (TUI) by combining physical objects 
and digital overlays \cite{Billinghurst2005}. Billinghurst et al. \cite{Billinghurst2005} noted that
tangible AR is exciting for quick modifications and real-time previews among expert designers and novices. However, the ergonomics of the hardware, like head-mounted displays, offered challenges for prolonged usage. Further parallels can be found in "Augmented Foam" \cite{Lee2005}, which allows a physical foam block to be treated as a canvas for hands-on sculpture and prototyping iteration. 

\subsubsection{Graspable User Interfaces:}
As discussed by Coutrix and Nigay \cite{Coutrix2006}, also discussed in Lee and Park \cite{Lee2005}, "graspable" interfaces track spatial and gesture input from multiple mixed tools, which can be used simultaneously. Their prototype RAZZLE explored design strategies to maintain continuity of interaction and minimize distraction shifts. Although it has been primarily demonstrated in a gaming context, the principles of spatial awareness and multi-tool feedback underpinning this research can help further augment future AR CAD systems. 
\subsubsection{Augmented Reality World Editing:}
Guida and Sra\cite{Guida2020} presented a mobile AR world editor that uses video inpainting to allow users to "erase" real-world objects in real time. Perhaps with less of an eye toward entertainment,
systems such as this one could aid product design by allowing engineers to visualize modifications of physical environments without actually building or moving anything. 
\subsubsection{Augmented Foam:}
Lee and Park \cite{Lee2005} introduced Augmented Foam, which combines AR technology with physical foam prototypes. By overlaying 3D virtual models onto a CNC-machined block, designers can use digital and physical references to expedite the iteration process. While promising, issues like handling accurate hand occlusion illustrate the absolute need for accurate tracking and calibration of interfaces in AR. 
\subsubsection{Product Behavior Simulation in Mixed Reality:}
Bruno et al. \cite{Bruno2010}, \cite{Harlan2023} examined mixed-reality prototyping that utilizes legitimate engineering data to simulate the behavior of products, including how a mechanical or electrical appliance reacts when the user interacts with it. Close integration with existing engineering tools reduces development overhead and creates more realistic test scenarios—a necessity for electromechanical devices, which inherently combine software, hardware, and user interfaces. 
\subsubsection{Hybrid AR Workstations:}
Hybrid AR workstations are interfaced between traditional CAD environment settings (for example, a powerful computer station) with visualization immersion setups (for example, headset or holographic displays) \cite{Harlan2023,Wang2006,PTC_AR}. Harlan et al. \cite{Harlan2023} recognized the foundation CAD activities-such as drafting, modeling, and early embodiment tasks, and developed AR modes for each. While geometry synchronization and UI design remain open problems, early studies suggest hybrid setups can streamline design workflows by allowing users to switch back and forth seamlessly between fine editing on a monitor and intuitive 3D manipulation in AR.
\subsubsection{Mixed Reality for Rapid Prototyping:}
Yin et al. \cite{Yin2023} presented a mixed-reality method for creating articulated prototypes, integrating hierarchical shape manipulation with real-time kinematic previews. Likewise, Mourtzis et al. \cite{Mourtzis2021} explored a cloud-based MR design framework, allowing engineers to co-visualize new parts with actual manufacturing systems. Since both of these approaches have difficulties in their alignment and measuring phases, they reflect the increasing trend where MR is integrated with parametric CAD/CAM workflows.

\subsection{VR Sketching Tools for Early-Stage Ideation}
Outside the center of AR-CAD, VR-sketching applications such as Gravity Sketch have matured and become popular with conceptual design and product styling. On the contrary, VR offers designers a chance to "walk inside" their sketches at a 1:1 scale. According to Joundi et al. \cite{Joundi2020}, this immersion was conducive to fast shape exploration; however, participants may have produced fewer concepts altogether than would have occurred with traditional sketching. This might have been because VR is more challenging to learn.

Ekströmer et al. \cite{Ekstromer2018} noted that while VR sketching promotes engagement, some users focus too much on making perfect geometry or keep hitting "undo," which prevents divergent thinking that is so essential in early idea generation. Lin\cite{Lin2022} showed that with a realistic 3D surface in VR, the sketch becomes less ambiguous and encourages participants to refine every concept heavily rather than produce several variations. Vo\cite{Vo2022} combined Gravity Sketch and 3D printing in luminaire design, showing intense excitement among design students but little gain in creativity. These findings indicate the VR's promising encouragement, albeit with new friction points, including setting up the device, mastering the controllers, and potential challenges with reinterpretation.

\section{AR-CAD System Description}
We present AR-CAD, which is an Augmented Reality (AR) based Computer-Aided Design (CAD) tool that we designed and developed to bridge the gap between real environment and 3D CAD.
\subsection{Hardware and Software Setup}
We developed AR-CAD for the Meta Quest 3 headset (but not limited to), equipped with four cameras, a depth sensor, and pass-through AR functionality as well as full VR capability. We used the Meta All-in-One SDK v64.0.0 \cite{MetaXRSDK} in Unity (2022.3.25f1) to use all of the features available in Quest 3 in AR-CAD. Using this SDK, we developed our basic features and functionalities. i.e hand/controller tracking and object manipulation. We used Unity’s Depth API \cite{OculusDepthAPI} to enable occlusion and mesh overlay effects, while an STL export mechanism was implemented using an open-source library \cite{WengSTL}. 


\subsection{Key Features of AR-CAD}
The key features that AR-CAD offers are as follows:
\subsubsection{Primitive Creation with Custom Dimensions:} Users can spawn parametric primitives (cubes, spheres, cylinders) and move the controllers (in 3 dimensions) while spawning to give a custom shape to the primitive and then joining shapes to combine geometry (a simple grouping, not a true boolean union). This enables constructing basic shapes in the real-world environment leveraging Augmented Reality.

\subsubsection{Manipulation of 3D Models:} Users can either grab the existing 3D objects using controllers by pressing the button on their middle finger on both controllers and then move their hands away or close to increase or decrease the size respectively. Or they can grab the 3D object with both hands, just like you naturally grab something, and then moving their hands away or close to change the size of that object. This direct manipulation is designed to mimic real-world object handling, providing a tangible feel.

\subsubsection{Mesh Creation from Plotted Points:} The most unique feature that AR-CAD offers is the creation of mesh from points, which allows the create points in 3D space and then generate a mesh surface through those points, using a library built from convex hull \cite{MIConvexHull} or an experimental concave hull we adapted from \cite{Moreira2007}. ConvexHull generates mesh facing outwards(round and smooth) between points while ConcaveHull generates inwards-facing (but straight in our case) between the points. This feature is unique in enabling free-form surface sketching, which would be cumbersome in traditional CAD without switching to complex
spline tools. This feature requires at least 3 points to create mesh as these algorithms create a triangular mesh, which needs at
at least 3 points to start.

\subsubsection{Flexible User Interface (UI):} 
\begin{figure}
\centering\includegraphics[width=\linewidth, alt = {AR-CAD UI}]{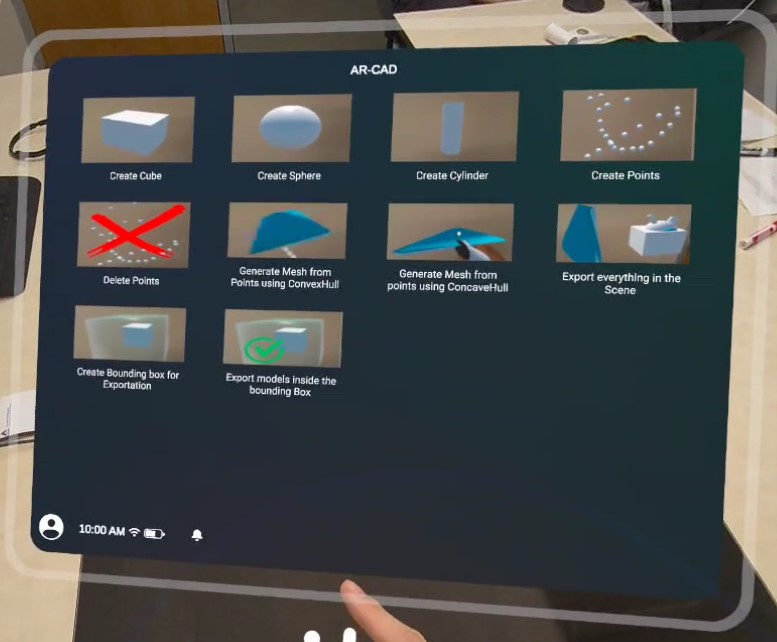}
\caption{AR-CAD Adjustable User Interface (UI)}\label{fig:1}
\end{figure}

The AR-CAD's UI is presented as a floating panel that users can reposition or resize for comfort. The panel includes buttons for shape creation, STL exportation, and adjusting the size of the UI boundary (highlighted as semi-transparent line at the boundary as shown in Fig.~\ref{fig:1}. Users can simply select any feature by clicking the index finger button on the controller or touching it with their finger.

\subsubsection{Export to STL:} At any time, users can export their models inside AR-CAD into STL format. They can decide if they want to export everything in the scene or a selective mesh. For everything in the scene, there is a button in UI, \textbf{\textit{Export everything in the Scene}} which exports mesh in STL format with name \textbf{exportedShape.stl}. For selective mesh, they can create a bounding box around the mesh by pressing, \textbf{\textit{Create a bounding box}} and then pressing \textbf{\textit{Export mesh inside bounding box}} to export it. There is an iterative counter on the exported objects and it will keep saving new exportations by changing the suffix \textbf{X} in \textbf{exportedShape\textit{X}.stl}. Once exported, you can get the STL file and use it as you want. This enables transitioning the AR-created models into a traditional CAD interface for further editing, thus supporting a hybrid workflow.

Together, these features provide an immersive design environment with basic CAD functionality (primitive modeling, spatial transformations, and model export). The intent was not to recreate every feature of a CAD modeling environment, but instead offer core tools that allows users to construct and assemble simple models entirely in AR. This system’s unique capabilities (like in-situ point-to-mesh creation and direct hand-based manipulation) aim to make 3D modeling more accessible and intuitive, especially for users without prior CAD training in any tool.
\section{Methodology}
We conducted a mixed-method, comparative user study to evaluate AR-CAD against a traditional CAD software called SolidWorks. The study design and procedures were approved by the University’s Research Ethics Board (Ethics ID: Pro00148144) and followed the guidelines outlined in our approved study plan. Both quantitative data (performance metrics, questionnaires) and qualitative data (observations, think-aloud transcripts, interviews) were collected for comprehensive analysis.

\subsection{Participants and Recruitment}
A total of 20 participants (17 male, 3 female; ages 18–49) were recruited from university students and staff. Email invitations were distributed to students, research associates, postdocs, and faculty in various engineering departments. The majority (17/20) had prior experience with some form of CAD software (e.g., SolidWorks, AutoCAD, Fusion 360), with self-rated CAD skills ranging from 4–9 out of 10. Half of the participants (10 out of 20) identified as CAD experts (self-rating$\geq$7), six participants as mediocre users (rating 4-6), one participant as a beginner (rating 1-3), and three participants (P3, P15, and P17) reported no prior CAD experience (we refer to them as CAD novices). In contrast, only about half had any prior exposure to AR/VR: 9 participants had tried VR or AR applications (mostly games or simple demos), but none had used an AR-based CAD tool before. Participants were thus generally familiar with traditional CAD to varying degrees, but all were new to AR-CAD. No compensation was provided beyond snacks, and informed consent was taken.
To remove bias, we randomly assigned participants to one of two groups:
\begin{itemize}
    \item \textbf{Group A:} Used Traditional CAD Interface first, then AR-CAD.
    \item \textbf{Group B:} Used AR-CAD first, then Traditional CAD Interface.
\end{itemize}

\subsection{Study Design}
We conducted our study in a suitable real environment where participants could solve design problems. We used AR-CAD on Quest 3 while the traditional CAD software (Dassault SolidWorks desktop version 2023) was used as the baseline, running on a standard workstation with a 32-inch 2D monitor, keyboard, and mouse. Both tools were used to perform equivalent modeling tasks for a fair comparison.
We used a within-subjects design: each participant performed modeling tasks using traditional CAD software and AR-CAD. To mitigate learning-order effects, the first tool was counterbalanced: 10 participants used AR-CAD first, then the traditional CAD interface and 10 did the reverse. For AR-CAD, a short video tutorial (3 minutes long) was given on using the AR interface in Quest 3 and how to use the controllers for AR-CAD.

Each participant completed two modeling tasks in each environment. We also asked participants to think aloud (verbalizing their thoughts, frustrations, or strategies) while designing in AR-CAD and the traditional CAD interface so we could record their thoughts and design actions. An experimenter was present to record times and observations; the AR sessions were recorded (with audio) via the headset, and traditional CAD interface sessions were screen recorded (with audio) using the sniping tool for later transcription of think-aloud recordings. Tasks were the same in both software; however, the approach to completing the tasks might have varied due to differences in the 3D and 2D interfaces.

\subsubsection{Tasks:}
Participants completed two design tasks in both tools as follows:
\begin{itemize}
\item \textbf{Single Part Creation:}  Participants were asked to create a simple shape of any dimensions (a simple 3D cube, sphere, or cylinder) in both tools according to their assigned group. A way to do it was by drawing or placing a rectangular profile and extruding it (in the traditional CAD interface) or spawning and scaling a cube (in AR-CAD). It tests basic shape creation and dimension control.  

\item \textbf{Assembly Construction:} If the participants were successful in task 1, they were asked to create any sort of structure that could connect two tables present in the room. To measure the distance between tables, they were provided with a measuring tape. In the traditional CAD interface, they needed to enter the exact distance to solve this problem but in AR-CAD, participants realized that they didn't need a value and just adjusted the size of the structure with their hands. This task tests spatial reasoning, multi-part placement, and the ability to align components. 
\end{itemize}

Participants were given a 10 to 15-minute limit per task to mimic time-constrained design scenarios. After completing both tasks in one platform, they filled out the System Usability Scale (SUS) \cite{Brooke1996} questionnaire and NASA Task Load Index (TLX) \cite{Hart1988} workload ratings for that tool. Then they repeated the tasks in the other tool, followed by the same questionnaires. Finally, a brief semi-structured interview was conducted to gather overall preferences, tool comparisons, and suggestions.

\subsubsection{Pilot Study:}
We conducted a pilot study with three participants, which helped us refine our study design. Participants reported issues with AR-CAD’s user interface, including difficulty pressing buttons. Therefore, we redesigned the UI to make it easier to use and added more flexibility so participants could adjust it to their preferences. We also realized that most of the participants who signed up had no experience with augmented/virtual reality interfaces, and those who did had never used an AR-based CAD tool. As a result, we created a three-minute tutorial video on using the AR interface in Quest 3 and how to use the controllers for AR-CAD.

\subsection{Measures and Data Collection}
While collecting data, we asked various questions related to participants' age, gender, level of education, field of study, department, current role, previous experience with CAD tools, or if they have any experience related to AR/VR tools with or without CAD. This helped us to see patterns and find correlations between different entities by making a correlation matrix. We removed the outliers later while analyzing the data.

\subsubsection{Performance Metrics:} We recorded the completion time for each task on each platform (if a participant did not finish and gave up, that was noted as a non-completion for that task). We also noted whether the participant successfully completed the task (binary for each task). A combined measure of overall success (both tasks completed vs not) per platform was later derived. The output models were saved for inspection of accuracy (though for our analysis we focused on completion and time, since precise quantitative accuracy measures were difficult given differences in every participant's approach).
\subsubsection{Usability and User Experience:} After using each tool, participants completed the standard SUS questionnaire, which yields a score from 0 to 100. We then used \cite{Brooke1996}'s formula to calculate the overall usability score for each participant. Additionally, we asked in the interview which tool they preferred and why.

\subsubsection{Cognitive Load:} We used the NASA-TLX after each participants performed tasks in each tool. Participants rated the mental demand, physical demand, temporal demand, effort, frustration, and perceived performance on a 10-point scales (0=Low, 10=High) for each tool. We used the average of the six TLX subscales as an overall workload score (unweighted TLX).

\subsubsection{Qualitative Observations:} The experimenter took notes during each session about any observable difficulties, strategies, or behaviors (e.g., “Participant is searching through menus”, “Participant moved around to view the model from different angles”). The think-aloud protocol and the recorded sessions of AR-CAD and the traditional CAD interface provided rich qualitative data on user thought processes and reactions. 

\subsubsection{Data Analysis Approach:} For quantitative data, we used paired statistical tests \cite{Ross2017} to compare AR-CAD and the traditional CAD interface on each metric (within-subject comparisons). We also explored whether prior experience (novice vs experienced) or tool order influenced outcomes by segmenting the data. For the scope of this paper, we are not writing the findings from our qualitative data and thematic analysis.

\section{Results and discussion}
We present the findings of our study by analyzing performance metrics, SUS, and NASA-TLX comparisons. These results provide a holistic comparison between AR-CAD and the traditional CAD interface. Where applicable, we report test statistics and p-values.

\subsection{Task Completion and Efficiency (RQ1 \& RQ2)} 
\begin{figure}[H]
\centering\includegraphics[width=\linewidth, alt = {Tasks completed vs CAD Expertise}]{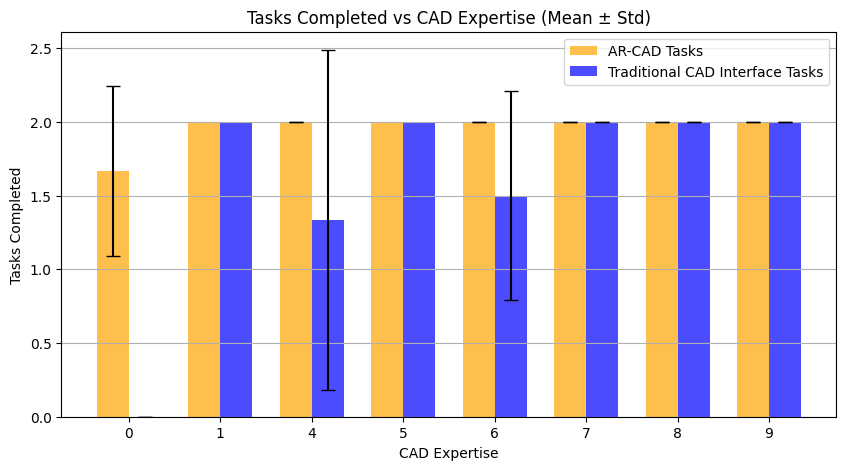}
\caption{Mean Number of tasks completed by participants Varying across cad expertise }\label{fig:2}
\end{figure}

All 20 participants were able to complete the required modeling tasks using the AR-CAD system, whereas several encountered difficulties in the traditional CAD interface. In fact, 19 out of 20 participants (95\%) successfully finished both Task 1 and Task 2 in AR-CAD within the time limits. In contrast, only 15 out of 20 (75\%) managed to fully complete both tasks in the traditional CAD interface. Four participants (including two of the three CAD novices) failed to produce a valid result in the traditional CAD interface and gave up due to confusion, even though the same individuals achieved the goals in AR-CAD. For example, participant P17 (a novice) struggled with the traditional CAD interface, spending ~8 minutes exploring menus without creating the correct model (ultimately reporting frustration and stopping), yet later in AR was able to methodically create and position shapes to form the bridge, completing the task. Similarly, P12 (some 2D drafting experience but new to parametric 3D CAD) got stuck with the sketch-extrude workflow in the traditional CAD interface and gave up, whereas in AR-CAD he successfully built the required structure after a brief learning period. This pattern suggests that AR-CAD offers a gentler learning curve, allowing users to achieve basic design outcomes more readily (addressing RQ1 for less experienced users as shown in Fig.~\ref{fig:2}). 

\begin{figure}[H]
\centering\includegraphics[width=\linewidth, alt = {Time vs expertise}]{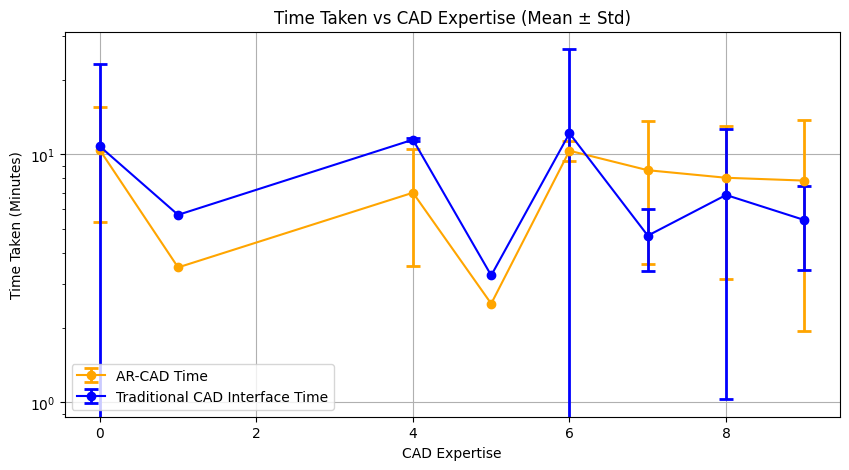}
\caption{Mean Time taken by participants varying across CAD expertise}\label{fig:3}
\end{figure}
Overall, in AR-CAD (39/40) 97.5\% tasks were completed while in the traditional CAD interface, only (31/40) 77.5\% tasks were completed which shows high success rate for AR-CAD. For the experienced CAD users, completion rates were high in both tools (most finished all tasks in both). However, even some experts noted that certain simple operations felt “quicker to just do directly” in AR. Conversely, a few participants observed that extremely precise adjustments (e.g., entering exact dimensions) were more straightforward in the traditional CAD interface due to its precise numeric input and snapping features – AR-CAD lacked a comparable precision input mechanism in this prototype, which could limit its effectiveness for fine-detail design refinement.

In terms of task efficiency (time to complete tasks, RQ2), the overall times were comparable between AR-CAD and the traditional CAD interface as shown in Fig.~\ref{fig:3}. The mean total time for Task 1 + Task 2 in AR-CAD was approximately 8.06 minutes (±4.24 minutes), while in the traditional CAD interface, it was approximately 7.85 minutes (±6.51 minutes). Although the mean time for AR-CAD is slightly higher, but the greater standard deviation of the traditional CAD Interface depicts higher variability. This small difference was not statistically significant (paired t(19) = 0.14, p = 0.89). In other words, on average participants took essentially the same amount of time in AR as in the traditional CAD environment to complete the tasks as shown in Fig.~\ref{fig:3}. It's also worth considering that experienced designers had years of experience in traditional CAD interfaces to achieve a high success rate in adequate time while novices were able to achieve similar results in AR-CAD without significant effort and in comparable time. 

\begin{figure}
\centering\includegraphics[width=\linewidth, alt = {Time vs participants}]{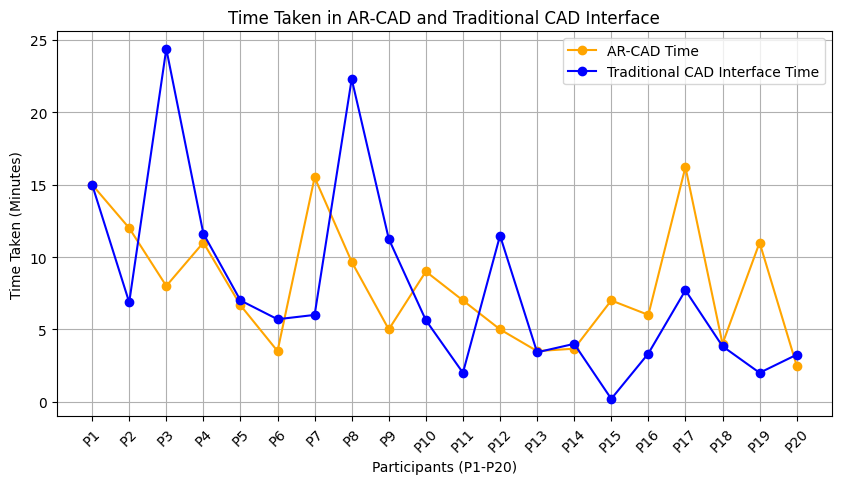}
\caption{Time taken by all participants to complete tasks}\label{fig:4}
\end{figure}

Individual performance varied widely according to experience and approach, as is apparent in Fig.~\ref{fig:4}. 
Some experienced traditional CAD interface users were extremely fast in that environment – for instance, P2 (an expert) completed the traditional CAD interface tasks in ~6.9 minutes total, whereas in AR-CAD he took ~12 minutes, as he spent extra time familiarizing himself with the new interface and even experimenting with additional features out of curiosity. On the other hand, some users (notably the novices and a few intermediates) were faster in AR-CAD than in the traditional CAD interface, since they struggled to remember the correct sequence of operations in the traditional CAD interface. For example, P8 (intermediate skill) took over 22 minutes in the traditional CAD interface (exceeding the time limit on Task 2) due to difficulty finding the right commands, but only ~9.5 minutes in AR-CAD by intuitively moving and resizing 3D objects he made. These extremes balanced out, resulting in nearly equal mean times. Another factor that was involved but not apparent in the graphs is the time participants spent playing with AR-CAD's feature out of curiosity even when they were finished with the tasks. Also in the traditional CAD interface, few participants could not complete tasks and gave up very early. An example of this is P15 (a novice), which can be seen in Fig.~\ref{fig:4}, who gave up on the traditional CAD interface just after 12s as he found the interface very cluttered and didn't want to continue. This increased the mean time to complete tasks in AR-CAD while it decreased for the traditional CAD interface even when the task completion rate was comparably low.

We also examined whether the order of tool use affected the times: participants who used AR-CAD first vs. second had no significant difference in the relative time performance. A two-sample test on the time differences confirmed no order effect (p > 0.5). Thus, AR-CAD’s efficiency was on par with the traditional CAD interface for the tested scenarios and significantly better in the case of novices due to a gentler learning curve, answering RQ1 in the affirmative: users can design a 3D model in AR-CAD about as effectively (in terms of time and success) as in the traditional CAD interface.
We also tested if the success rates differed significantly. Using a McNemar test \cite{PemburySmith2020} for the 19 vs.\ 15 completion counts showed a trend favoring AR (binomial \(p \approx 0.125\), not below 0.05), reflecting the higher completion in AR-CAD but the sample size was too small for significance.

\begin{figure}
\centering\includegraphics[width=\linewidth, alt = {sus vs expertise}]{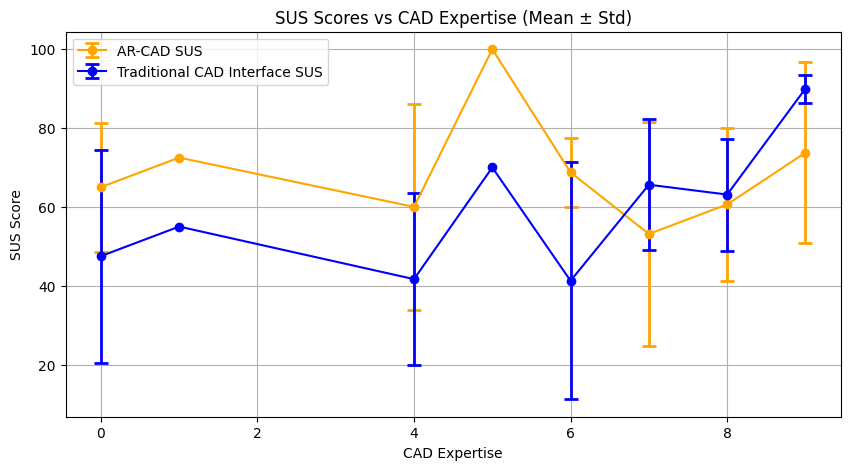}
\caption{Mean System Usability Scores (SUS) varying across cad expertise}\label{fig:5}
\end{figure}
\subsection{Usability and User Satisfaction (RQ3)} The System Usability Scale (SUS) results revealed a slight overall preference towards the AR-CAD system. AR-CAD received an average SUS score of 64.4 (±20.84), versus 58.5 (±21.76) for the traditional CAD interface (scores out of 100). For context, SUS scores around 50–60 are considered mediocre, while scores above ~68 are considered above-average usability\cite{Bangor2008}. So both tools in this study landed in the mid-range, with AR-CAD edging higher by ~6 points. However, this difference was not statistically significant (p(19) = 0.849, p = 0.41). The high variability in ratings across participants meant that we could not conclusively say one system is quantitatively more usable than the other. 

Looking deeper, the SUS results differed starkly by user experience level as shown in Fig.~\ref{fig:5}. The novice CAD users gave AR-CAD much higher usability ratings than the traditional CAD interface. The three novices (P3, P15, P17) rated AR-CAD SUS an average of ~65, versus ~47 for the traditional CAD interface – two of them in fact gave the traditional CAD interface extremely low scores (e.g., P17: SUS 25) due to their inability to effectively use it. One novice (P3) did rate the traditional CAD interface relatively high (77.5) despite failing the tasks, perhaps reflecting a perception that “it would be good if I knew how to use it” – an interesting anomaly. In general, novices found the traditional CAD interface very hard to use, bringing its average SUS down, whereas they found AR-CAD user-friendly. Meanwhile, experienced CAD users’ SUS scores varied. Many experienced participants rated the traditional CAD interface in the 70–85 range (indicating good usability for them, as they are familiar with it) and gave AR-CAD similarly high scores if they had a positive experience. However, a few experienced users gave AR-CAD low SUS despite completing tasks, citing the lack of advanced functionality. For example, P16 (highly experienced) scored the traditional CAD interface 85 but only gave AR-CAD 28, commenting that AR-CAD felt “too limited; I missed the detailed control I have in the traditional CAD interface.” On the other hand, P17 (novice) scored AR-CAD 67.5 vs the traditional CAD interface 25, aligning with her experience that AR-CAD was easy to use whereas the traditional CAD interface was overwhelming. Therefore, we also need to consider that experienced designers were less open to learning new tools as they had years of training on the traditional CAD interface before they became comfortable with making complex designs to complete tasks. At the same time, novices found AR-CAD useful without any hesitation as it made the learning curve shallow for them as they didn't have to struggle with complex designs during task 2 due to AR-CAD's features utilizing spatial visualization. 

Overall, 12 out of 20 participants had higher SUS for AR-CAD, 6 preferred the traditional CAD interface, and 2 rated them roughly equal. This indicates that a slight majority favored the usability of AR-CAD, which was driven mainly by those without extensive CAD background. This also shows the importance of AR-CAD as a tool for entering CAD for novices as highlighted by \cite{Zhang2025}, how AR modeling interface boosted user engagement for novices compared to traditional CAD tools.

\subsection{Cognitive Workload and Comfort (RQ3 continued)} 
\begin{figure}
\centering\includegraphics[width=\linewidth, alt = {tlx vs users}]{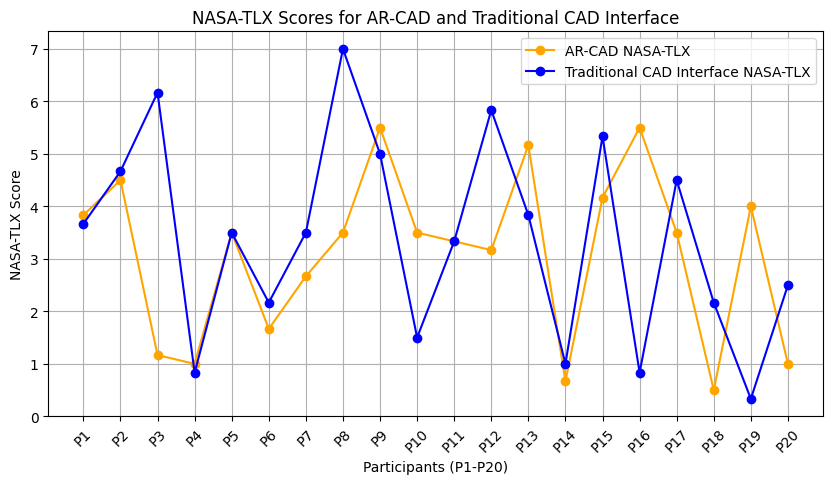}
\caption{NASA-TLX (Task Load Index) varying across Participants}\label{fig:6}
\end{figure}

\begin{figure}
\centering\includegraphics[width=\linewidth, alt = {tlx vs expertise}]{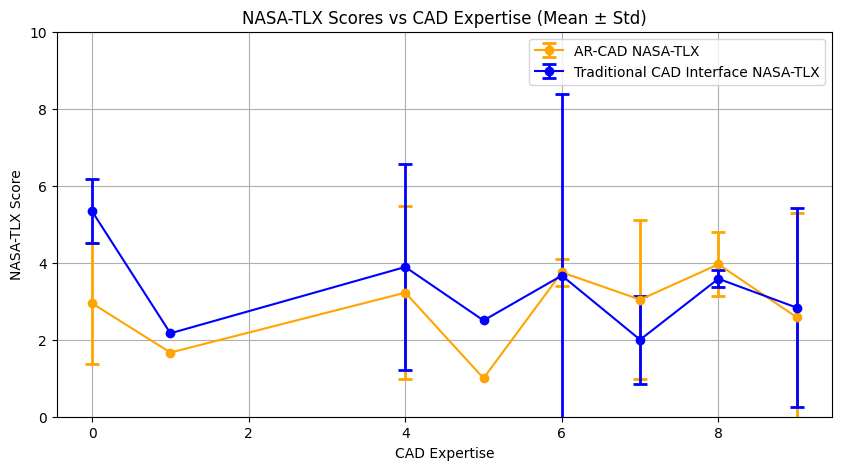}
\caption{Mean NASA-TLX (Task Load Index) varying across cad expertise}\label{fig:7}
\end{figure}

\begin{figure}
\centering\includegraphics[width=\linewidth, alt = {TLX sub components}]{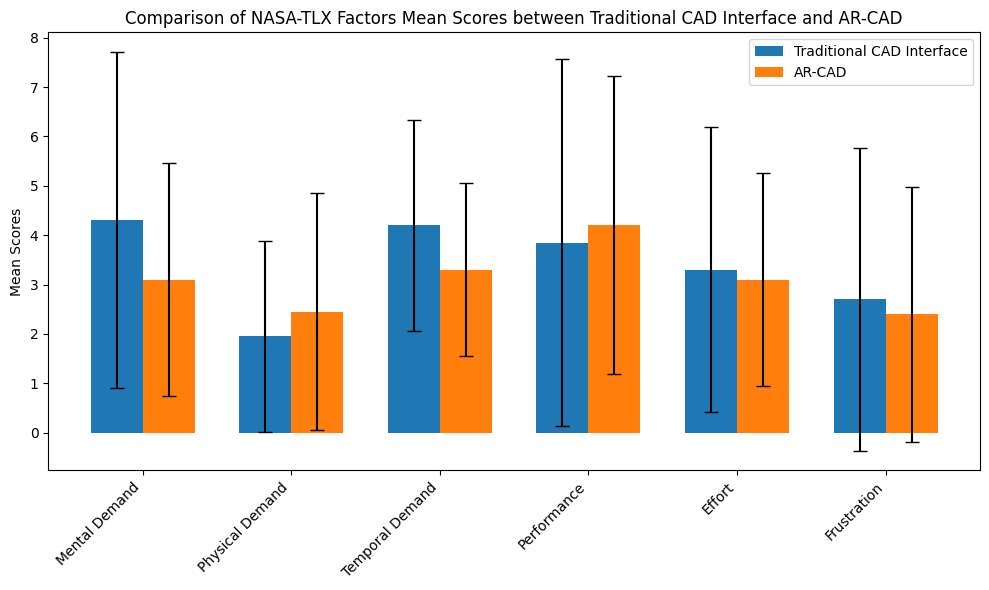}
\caption{Mean values of NASA-TLX sub-components}\label{fig:8}
\end{figure}

The NASA-TLX results showed no strong differentiation in overall cognitive load between AR-CAD and the traditional CAD interface. The average TLX overall score (on a 0--10 scale for each factor, unweighted average) was $\sim 3.092 \, (\pm 1.596)$ for AR-CAD and $\sim 3.383 \, (\pm 1.943)$ for the traditional CAD interface.
The slight difference favored AR-CAD (lower workload), but it was not statistically significant (p(19) = -0.591, p = 0.561). The results for NASA-TLX were different among participants and were not significantly related to their CAD expertise. Fig.~\ref{fig:6} shows the NASA-TLX varying across different participants and Fig.~\ref{fig:7} shows NASA-TLX varying across CAD expertise which depicts that it was slightly higher for AR-CAD in the case of experienced designers (showing that they felt more comfortable with traditional CAD interface due to their years of training in it) and for the traditional CAD interface, it was higher among novices due to its complexity and steeper learning curve. In plain terms, participants on average found both tools somewhat demanding but manageable, with no clear winner in the objective workload. However, examining specific TLX sub-components provides insight into different kinds of load in each tool as shown in Fig.~\ref{fig:8}. 
Notably, participants reported higher physical and performance demand (since it was their first time using it) in AR-CAD. Wearing the headset and holding up one’s arms to manipulate objects led to some fatigue. In contrast, the traditional CAD interface had virtually no physical demand (all users were seated using a mouse/keyboard). This likely made the Physical Demand TLX rating higher for AR-CAD and lower for the traditional CAD interface as shown in Fig.~\ref{fig:8}. On the other hand, mental demand and frustration were often higher in the traditional CAD interface, especially for those who got stuck. Participants P3 and P17 (both novices) described the traditional CAD interface experience as “mentally taxing – I was trying to recall which button to press and what steps to do.” This aligns with the idea that the complex, abstract operations in a traditional CAD tool (sketch, constrain, extrude, etc.) impose a higher cognitive load on newcomers. In AR-CAD, even though these users had never seen the system before, they did not report the same level of mental strain. P3 noted, “I could figure things out by just looking and trying; it made sense visually,” indicating that the spatial, direct manipulation interface helped reduce the mental overhead of mapping out 2D plans or remembering sequences. We observed that temporal demand (time pressure) had mixed responses: some felt more rushed in the traditional CAD interface because they were conscious of time ticking while hunting through menus, while others felt more pressure in AR-CAD because the novelty made them slower initially. The effort and frustration ratings strongly reflected whether a participant encountered a roadblock. Those who gave up on the traditional CAD interface had frustration scores near the maximum for that condition, skewing the traditional CAD interface average higher. In contrast, AR-CAD frustration ratings were generally low; even when users faced minor technical issues, they tended to still feel they made progress.

Overall, while the quantitative workload scores were similar, these nuances show how the type of load differed: AR-CAD introduced some physical effort but eased the cognitive puzzle of modeling, whereas the traditional CAD interface was mentally demanding especially for the novices, though physically effortless. No participant reported severe discomfort using AR-CAD—no cases of motion sickness or serious dizziness were noted (likely because the Quest 3’s pass-through AR keeps one grounded in the real world, and our sessions were relatively short). A few mentioned the headset’s weight and the mild strain of holding arms up, but said it was acceptable for the task duration. Meanwhile, working on a PC was second nature for all, with no physical complaints. In summary, AR-CAD slightly reduced overall workload relative to traditional CAD; it appears to trade off higher physical effort for lower mental effort. This addresses RQ3 by indicating that a well-designed AR interface can be adopted without overloading the user, and may even alleviate certain cognitive burdens.

In aggregate, our quantitative analysis finds that AR-CAD is competitive with traditional CAD software in accomplishing standard modeling tasks. Task success rates were higher in AR-CAD, especially for novices, though experienced users performed well in both. Task completion times were statistically equivalent between AR-CAD and the traditional CAD Interface. Perceived usability was slightly higher for AR-CAD on average (driven by strong novice preference), but overall usability differences were not significant due to mixed and biased opinions from experienced designers. Cognitive load was similar between the two, with AR perhaps reducing mental effort at the cost of some physical effort. These quantitative results suggest that AR-CAD can achieve comparable efficiency and workload to mature traditional CAD systems, even for the complex modeling tasks we tested.

\section{Conclusion}
In this paper, we conducted a mixed-method comparative evaluation of an Augmented Reality CAD system (AR-CAD) versus a conventional CAD tool (SolidWorks). Our results demonstrated that AR-CAD could substantially improve efficiency, enabling participants to complete complex tasks faster by utilizing its spatial visualization than in the traditional CAD interface, especially for novices, while maintaining similar accuracy levels.

Novice users found the AR-CAD interface more intuitive and immersive, especially for complex designs, while experienced designers still preferred the traditional CAD interface for certain precision and advanced features. Overall, AR-CAD is a promising tool, especially in the early design stages or when a strong spatial context is advantageous. Given its limited features and being an early-stage prototype, it still performs well with slight limitations on precision. It could replace traditional CAD interfaces in the future, depending on advancements in technology and its features, and has proved to be a better tool for novices entering CAD modeling.

\section{Limitations and Future Work}
Certain limitations in this study could be improved in future works that could also explore certain avenues discussed below:

\textbf{Refine AR-CAD Interface:} Future studies could explore user feedback and engagement with AR-based CAD tools that incorporate advanced features such as smart dimensions, boss extrude, and cut extrude to assess whether experienced designers perceive them as superior alternatives. A limitation of AR-CAD was the high score for physical demand. As AR/VR hardware continues to evolve, future studies could explore whether users feel better using AR-based CAD tools and whether or not their reports of physical demand change. Experienced users also mentioned that AR-CAD's current prototype lacks precise modeling tools, which is another area for future research.

\textbf{Larger-Scale User Studies:} Future studies could engage more participants with diverse experience levels to examine how
performance evolves over long-term AR-CAD usage and how it affects users. Future work can also target groups with specific CAD expertise and thorough user experience study significantly affects how it could be improved for a particular group. Future studies could also explore measuring the precise time for task completion, as many participants in this study spent time playing with the features even when they were done with the tasks which increased the mean completion time for AR-CAD.

\textbf{Industrial Usage:} The main advantage of an AR-based CAD system is its visualization, rapid prototyping and in-scene modeling. Unfortunately, we did not have such tasks that heavily relied on visualization, especially in an industrial setup where AR-CAD's visualization capabilities could be used to validate the 3D designs in a real environment. Future studies could set up experiments where designing could be highly affected by visualization and in-scene modeling to test AR-CAD's true potential.

\textbf{Hybrid Workflows:} As shown in our study and literature, Hybrid workflows might be better using AR’s visualization capabilities and advanced features of traditional CAD. Future studies
could explore how such type of systems could be designed and integrated in a way that 3D modeling becomes fast, accurate and
convenient. 

\textbf{Sustainability:} Conventional methods involve a lengthy loop where measurements are taken before prototyping and then manufacturing 3D parts, where slight errors can lead to part wastage and, ultimately, significant material loss. Future studies could explore how reliable AR-CAD's in-scene modeling is without the need for measurements, and how sustainable the entire loop becomes.

\section{Acknowledgments}
We thank all pilot and study participants for their time and feedback. Ethics approval was obtained from the University of Alberta Research Ethics Board (Pro00148144). The authors acknowledge research funding from the University of Alberta Faculty of Engineering’s Research Exploration Fund. The authors also acknowledge the Unity and Meta Quest development teams for providing foundational AR/VR tools used in developing AR-CAD.

\nocite{Buchner2022,Tiwari2023,Arslan2024} 
\bibliographystyle{asmeconf}

\end{document}